# Giant magnetic field from moiré induced Berry phase in homobilayer semiconductors

Hongyi Yu[1*], Mingxing Chen[2], Wang Yao[1*]

[1] Department of Physics and Center of Theoretical and Computational Physics, University of Hong Kong, Hong Kong, China

[2] School of Physics and Electronics, Hunan Normal University, Changsha, Hunan 410081, China

*Correspondence to: yuhongyi@hku.hk, wangyao@hku.hk

**When quasiparticles move in condensed matters, the texture of their internal quantum structure as a function of position and momentum can give rise to Berry phases that have profound effects on the material's properties. Seminal examples include the anomalous Hall and spin Hall effects from the momentum-space Berry phases in homogeneous crystals. Here, we explore a conjugate form of electron Berry phase arising from the moiré pattern: the texture of atomic configurations in real space. In homobilayer transition metal dichalcogenides, we show the real-space Berry phase from moiré patterns manifests as a periodic magnetic field with magnitudes up to hundreds of Tesla. This quantity distinguishes moiré patterns from different origins, which can have identical potential landscape but opposite quantized magnetic flux per supercell. For low energy carriers, the homobilayer moirés realize topological flux lattices for the quantum spin Hall effect. An interlayer bias can continuously tune the spatial profile of moiré magnetic field, whereas the flux per supercell is a topological quantity that can only have a quantized jump observable at moderate bias. We also reveal the important role of the non-Abelian Berry phase in shaping the energy landscape in small moiré patterns. Our work points to new possibilities to access ultra-high magnetic fields that can be tailored on the nanoscale by electrical and mechanical controls.**

## Introduction

In van der Waals layered structures, the creation of long period moiré patterns (via a small lattice mismatch between the layers) has become a powerful method for engineering superlattice electronic, optical and topological properties. Experiments have discovered emergent electron phenomena from graphene moiré superlattices, including the fractal quantum Hall effect(1-3), gate-tunable Mott insulators(4, 5) and superconductivity(6-11). In heterostructures of 2D semiconductors, the moiré pattern leads to nanoscale patterning in the energy and emission features of optical excitations and thereby significantly changes the heterostructure optical responses(12-17). Furthermore, theories have predicted the emergence of novel topological insulating states in transition metal dichalcogenides (TMD) heterobilayer moiré(18) and twisted homobilayers(19).

The nature of the moiré pattern as a spatial texture of atomic configurations suggests that the Berry phase effect can be an indispensable part of moiré superlattice physics. In a smoothly varying crystalline environment like a long-period moiré, the Berry phase has several important manifestations in the electron's equations of motion(20): $\dot{\mathbf{R}} = \frac{\partial E}{\partial \mathbf{k}} - \dot{\mathbf{k}} \times \mathbf{\Omega}^k$, $\dot{\mathbf{k}} = -\frac{\partial E}{\partial \mathbf{R}} + \dot{\mathbf{R}} \times \mathbf{\Omega}^R$, where $E$, $\mathbf{R}$ and $\mathbf{k}$ are the energy, position and momentum of an electron wavepacket. $\mathbf{\Omega}^k \equiv i \langle \frac{\partial u}{\partial \mathbf{k}} | \times | \frac{\partial u}{\partial \mathbf{k}} \rangle$ and $\mathbf{\Omega}^R \equiv i \langle \frac{\partial u}{\partial \mathbf{R}} | \times | \frac{\partial u}{\partial \mathbf{R}} \rangle$ are Berry curvatures from the momentum-space and real-space textures respectively in the electron's spin and pseudospin wavefunction $u$. Homogeneous crystals can only have the momentum space curvature $\mathbf{\Omega}^k$, and the resultant anomalous velocity $-\dot{\mathbf{k}} \times \mathbf{\Omega}^k$ is responsible for the various Hall effects(20). Spatial inhomogeneity can give rise to the real-space curvature $\mathbf{\Omega}^R$ that plays the same role as a magnetic field. Such an emergent magnetic field also generates a Hall current, which has drawn remarkable interest in explorations of magnetization skyrmions and domains(21-23).

Here, we show that the real-space Berry phase from moiré patterns realizes a nanoscale patterned magnetic field for the massive Dirac fermions in TMD homobilayers. The field is normal to the plane, with the full symmetry of the moiré pattern. The magnetic flux per moiré supercell has a quantized value, while its sign distinguishes moirés introduced by a uniaxial strain from those by a twisting or biaxial strain, although they have the same potential landscape. For low energy carriers, the complex hopping in the moiré potential as determined by the Berry phase realizes the topological flux lattices hosting quantum spin Hall effect. Strain control of the moiré period ($b$) can dramatically tune the moiré magnetic field with the $b^{-2}$ scaling, while preserving the flux per supercell. The field magnitude can reach hundreds of Tesla at a moiré size of $b \sim$ **10** nm. The profile of the moiré magnetic field can also be electrically tuned through an interlayer bias, and a topological transition occurs at a moderate bias where the magnetic flux per supercell has a quantized jump. We also reveal that the geometric scalar potential due to the non-Abelian nature of Berry phase can qualitatively change the potential landscape in a small moiré.

## Results

TMD monolayers have their spin-valley locked conduction and valence band edges at the $\pm \mathbf{K}$ corners of Brillouin zone, which are described by the massive Dirac model(24). We focus here on the physics at these band edges in a homobilayer moiré, formed by a small twisting or strain applied to one layer in R-stacking(25) (c.f. Fig. 1b and Table 1). The interlayer hopping of the valley electron is sensitive to the atomic registry between the layers(18, 26). The smooth variation of the interlayer registry in a long-period moiré thus leads to the dependence of the hopping on the position $\mathbf{R}$ (27, 28). In the basis $\{|t\rangle_c, |b\rangle_c, |t\rangle_v, |b\rangle_v\}$, denoting conduction (*c*) and valence (*v*) band-edge states from top (*t*) and bottom (*b*) layers respectively, the Hamiltonian for the two coupled massive Dirac cones reads,

$$\begin{pmatrix} E_g/2 + \hat{H}_c(\mathbf{R}) & i\hbar v_D \partial_+ \\ i\hbar v_D \partial_- & -E_g/2 + \hat{H}_v(\mathbf{R}) \end{pmatrix}. \qquad (1)$$

The 2x2 block $\hat{H}_v(\mathbf{R}) = \delta(\mathbf{R}) + \hat{\sigma}_z \Delta(\mathbf{R}) + \hat{\sigma}_+ h(\mathbf{R}) + \hat{\sigma}_- h^*(\mathbf{R})$ describes the spatially varying interlayer coupling effects in valence bands. $\hat{\boldsymbol{\sigma}}$ is the layer pseudospin. $h$ is the hopping matrix element between the valence band-edges from the two layers. The interlayer coupling also leads to energy shifts of the band-edges accounted by $\delta(\mathbf{R})$ and $\Delta(\mathbf{R})$ (19, 26). $\hat{H}_c$ describes similar interlayer coupling effects in the conduction bands. The $\hat{H}_c$ and $\hat{H}_v$ blocks are coupled through the $\partial_\pm \equiv \frac{\partial}{\partial X} \pm i \frac{\partial}{\partial Y}$ term of the monolayer Dirac Hamiltonian. Eq. (1) is for the $-\mathbf{K}$ valley, which has spin-up states only at the band edge, and the $\mathbf{K}$ valley is its time reversal.

**Non-Abelian Berry connection in real space.** We derive an effective Hamiltonian for holes in the bilayer moiré, whereas that for the electron is similar. It is convenient to go to the basis spanned by the eigenstates of interlayer coupling $\hat{H}_v(\mathbf{R})$: $|+\rangle = \hat{U}(\mathbf{R})|t\rangle_v$ and $|-\rangle = \hat{U}(\mathbf{R})|b\rangle_v$, where $\hat{E}_v(\mathbf{R}) \equiv \text{diag}(E_+(\mathbf{R}), E_-(\mathbf{R})) = \hat{U}(\mathbf{R})\hat{H}_v(\mathbf{R})\hat{U}^\dagger(\mathbf{R})$. Through a unitary transformation by $\hat{U}(\mathbf{R})$, the bilayer Hamiltonian in Eq. (1) becomes:

$$\hat{H}_{\text{moiré}} = \begin{pmatrix} \hat{U}(\mathbf{R})\hat{H}_c(\mathbf{R})\hat{U}^\dagger(\mathbf{R}) + E_g/2 & \hbar v_D \left(i\partial_+ + \mathbf{e}_+ \cdot \hat{\mathbf{A}}(\mathbf{R})\right) \\ \hbar v_D \left(i\partial_- + \mathbf{e}_- \cdot \hat{\mathbf{A}}(\mathbf{R})\right) & \hat{E}_v(\mathbf{R}) - E_g/2 \end{pmatrix}. \qquad (2)$$

In the off-diagonal block, the matrix $\hat{\mathbf{A}}(\mathbf{R})$ is the non-Abelian real-space Berry connection in the basis of the layer-pseudospin eigenstates:

$$\hat{\mathbf{A}}(\mathbf{R}) \equiv i\hat{U}(\mathbf{R})\frac{\partial \hat{U}^\dagger(\mathbf{R})}{\partial \mathbf{R}} = \begin{pmatrix} \mathbf{A}^{++} & \mathbf{A}^{+-} \\ \mathbf{A}^{-+} & \mathbf{A}^{--} \end{pmatrix}, \qquad (3)$$

where $\mathbf{A}^{\pm\pm} = i\langle \pm | \frac{\partial}{\partial \mathbf{R}} | \pm \rangle$, and $\mathbf{A}^{\pm\mp} = i\langle \pm | \frac{\partial}{\partial \mathbf{R}} | \mp \rangle$. $\mathbf{e}_\pm \equiv \mathbf{e}_x \pm i\mathbf{e}_y$, $\mathbf{e}_x$ and $\mathbf{e}_y$ are the unit vectors along *x* and *y* directions.

At the three high symmetry locales in the moiré (Fig. 1a-b), rotational symmetry dictates that $h$ vanishes at $B$ and $C$, while $\Delta$ vanishes at $A$ and takes opposite values at $B$ and $C$ (Fig. 2b). These determine a real-space texture in the two branches of layer pseudospin eigenstates. For example, in the $|+\rangle$ state, $\langle\hat{\sigma}\rangle$ is in-plane at $A$, and points out-of-plane in opposite directions at $B$ and $C$, as illustrated in Fig. 1b. Such a spatial texture of the layer pseudospin gives rise to a non-trivial Berry connection.

With the large gap $E_g$, the coupling to the conduction states can be perturbatively eliminated, and the hole Hamiltonian keeping its leading effects becomes,

$$\hat{H}_{\text{moiré}}^{(\text{hole})} \approx -\hat{E}_v(\mathbf{R}) + \frac{\hbar^2}{2m^*}\left(i\frac{\partial}{\partial \mathbf{R}} + \hat{\mathbf{A}}(\mathbf{R})\right)^2 \quad (4)$$

where the dropped terms are $O\left(\left(\frac{\hbar v_D}{E_g}\right)^3\right)$. $m^* \equiv \frac{E_g}{2v_D^2}$ is the Dirac mass.

**Periodic magnetic fields in different types of moiré.** We focus first on the diagonal Berry connection $\mathbf{A}^{\pm\pm}$. The main effect of off-diagonal connection $\mathbf{A}^{\pm\mp}$ is a correction to the scalar potential landscape that is negligible in a sufficiently large moiré, which we will revisit later. If $\mathbf{A}^{\pm\mp}$ is simply dropped in Eq. (4), the two branches of layer-pseudospin eigenstates are decoupled, each described by an effective Hamiltonian: $-E_\pm(\mathbf{R}) + \frac{\hbar^2}{2m^*}(i\nabla + \mathbf{A}^{\pm\pm})^2$. The moiré effects manifest as the scalar superlattice potential $-E_\pm(\mathbf{R})$, and more intriguingly the vector potential $\mathbf{A}^{\pm\pm}$ that generates a real-space magnetic field: $\boldsymbol{\mathcal{B}}^\pm = \left(\frac{\partial A_y^{\pm\pm}}{\partial X} - \frac{\partial A_x^{\pm\pm}}{\partial Y}\right)\mathbf{e}_z$.

We establish below the quantitative forms of the superlattice potential $E_\pm(\mathbf{R})$ and the periodic magnetic field $\boldsymbol{\mathcal{B}}^\pm$. In a long period moiré, the interlayer coupling parameters at location $\mathbf{R}$ are determined by the local atomic registry, which can be parameterized by the in-plane displacement $\mathbf{r}$ between a near-neighbor pair of atoms from opposite layers (Fig. 1, inset). Under the two-center-approximation and keeping only the leading Fourier components(26), we find the dependences of $\Delta$, $\delta$ and $h$ on the registry $\mathbf{r}$,

$$\Delta(\mathbf{r}) = \frac{\Delta_0}{9}\left|e^{i\mathbf{K}_1\cdot\mathbf{r}} + e^{i\left(\mathbf{K}_2\cdot\mathbf{r} - \frac{2\pi}{3}\right)} + e^{i\left(\mathbf{K}_3\cdot\mathbf{r} - \frac{4\pi}{3}\right)}\right|^2 - \frac{\Delta_0}{9}\left|e^{i\mathbf{K}_1\cdot\mathbf{r}} + e^{i\left(\mathbf{K}_2\cdot\mathbf{r} + \frac{2\pi}{3}\right)} + e^{i\left(\mathbf{K}_3\cdot\mathbf{r} + \frac{4\pi}{3}\right)}\right|^2,$$

$$\delta(\mathbf{r}) = -\frac{\delta_0}{9}\left|e^{i\mathbf{K}_1\cdot\mathbf{r}} + e^{i\left(\mathbf{K}_2\cdot\mathbf{r} - \frac{2\pi}{3}\right)} + e^{i\left(\mathbf{K}_3\cdot\mathbf{r} - \frac{4\pi}{3}\right)}\right|^2 - \frac{\delta_0}{9}\left|e^{i\mathbf{K}_1\cdot\mathbf{r}} + e^{i\left(\mathbf{K}_2\cdot\mathbf{r} + \frac{2\pi}{3}\right)} + e^{i\left(\mathbf{K}_3\cdot\mathbf{r} + \frac{4\pi}{3}\right)}\right|^2,$$

$$h(\mathbf{r}) = h_0\left(e^{i\mathbf{K}_1\cdot\mathbf{r}} + e^{i\mathbf{K}_2\cdot\mathbf{r}} + e^{i\mathbf{K}_3\cdot\mathbf{r}}\right) + h_1\left(e^{-2i\mathbf{K}_1\cdot\mathbf{r}} + e^{-2i\mathbf{K}_2\cdot\mathbf{r}} + e^{-2i\mathbf{K}_3\cdot\mathbf{r}}\right), \quad (5)$$

where $\Delta_0$, $\delta_0$, $h_0$ and $h_1$ are real constants. $\mathbf{K}_1 \equiv \mathbf{K} = (0, K)$, $\mathbf{K}_2 \equiv \hat{C}_3\mathbf{K}$, $\mathbf{K}_3 \equiv \hat{C}_3^2\mathbf{K}$ are respectively the three corners of the monolayer Brillouin zone related by $2\pi/3$-rotations. On the other hand, the split valence band edges $E_\pm = \delta \pm \sqrt{|h|^2 + \Delta^2}$ can be obtained from *ab initio* band structures of lattice-matched bilayers of various registry $\mathbf{r}$. In Fig. 2a, the *ab initio* calculated $E_\pm$ as functions of $\mathbf{r}$ are shown as the symbols, which are remarkably well-fitted by the solid curves from Eq. (5), with $\delta_0 =$ **1 meV**, $\Delta_0=$ **22.3 meV**, $h_0 =$ **7.1 meV** and $h_1 = -$**1.2 meV**.

The registry **r** is a function of position **R** in the moiré. The mapping **r(R)**, together with Eq. (5), gives the position dependences of the interlayer coupling parameters in $\hat{H}_v(\mathbf{R})$, i.e. $\Delta(\mathbf{R}) = \Delta(\mathbf{r}(\mathbf{R}))$ etc. The resultant spatial texture of layer-pseudospin eigenstates, the scalar potential $-E_{\pm}(\mathbf{R})$ and the moiré magnetic field $\mathcal{B}^{\pm}$ can then be determined. Table 1 summarizes the mapping function **r(R)**, for three different types of moiré pattern formed respectively by applying a twisting, a biaxial strain, and an area-conserving uniaxial strain to one layer. All three moiré patterns realize hexagonal superlattices, with the identical scalar potential landscape $-E_{\pm}(\mathbf{R})$ (except for a 90º rotation of the superlattice in the twisting case, c.f. Table 1).

Fig. 2d and 2e show the spatial profile of layer pseudospin vector $\langle \hat{\sigma} \rangle_+$ and a moiré magnetic field $\mathcal{B}^+$ in a moiré supercell in the lower energy hole branch. Remarkably, while the twisting moiré and biaxial moiré have identical pseudospin textures and magnetic field distribution, the uniaxial moiré is of a distinct texture that leads to an opposite magnetic field distribution. The magnetic flux per supercell $\int_{\text{SC}} \mathbf{e}_z \cdot \mathcal{B}^+ d\mathbf{R}$ is a quantized value: **2**$\pi$ for twisting moiré and biaxial moiré and $-$**2**$\pi$ for uniaxial moiré, corresponding to the fact that the pseudospin texture is of a skyrmion(19) and anti-skyrmion configuration respectively. Thus, the three different origins give rise to two topologically distinct types of moiré patterns. The moiré magnetic field $\mathcal{B}^+$ is peaked between the *B* and *C* points (Fig. 2d-e). In a moiré of period $b =$ **10** nm, this magnetic field reaches a peak value of 200 Tesla, comparable in size to the giant pseudo magnetic field from an inhomogeneous strain in a graphene nano-bubble(29-31).

**Topological flux lattice and quantum spin Hall effect.** The periodic scalar potential and magnetic field generated by the moiré together define a flux superlattice. Some remarkable features include: (1) the flux per supercell is quantized and independent of the moiré period; (2) the two layer-pseudospin branches $|+\rangle$ and $|-\rangle$ have opposite magnetic fields $\mathcal{B}^+ = -\mathcal{B}^-$, and distinct trapping locations in their potentials $-E_+(\mathbf{R})$ and $-E_-(\mathbf{R})$ (Fig. 2a); (3) spin up and down carriers experience opposite magnetic fields as required by the presence of time-reversal symmetry.

Holes can be trapped in the low energy pseudospin branch $|+\rangle$ at the three high symmetry locales *A*, *B* and *C* in a moiré supercell (c.f. Fig. 2c). Low energy holes hopping between these trapping sites can then be described by a three-orbital tight-binding model (Fig. 3a),

$$\hat{H}_{TB} = \sum_i \left( \varepsilon_A \hat{A}_l^\dagger \hat{A}_l + \varepsilon_B \hat{B}_l^\dagger \hat{B}_l + \varepsilon_C \hat{C}_l^\dagger \hat{C}_l \right) \qquad (6)$$
$$- \sum_{\langle l,m \rangle} \left( t_1 \textbf{exp}(i\phi_1^{l,m}) \hat{A}_l^\dagger \hat{B}_m + t_2 \textbf{exp}(i\phi_2^{l,m}) \hat{A}_l^\dagger \hat{C}_m + t_3 \hat{C}_l^\dagger \hat{B}_m + \textbf{h.c.} \right).$$

Here $\hat{A}_l$, $\hat{B}_l$ and $\hat{C}_l$ are annihilation operators for the hole trapped at $A$, $B$ and $C$ sites in $l$th supercell, and $\langle ... \rangle$ runs over nearest-neighbor pairs of sites. The sum of hopping phases $\phi_1$ and $\phi_2$ around any closed loop equals the magnetic flux enclosed. The equilateral triangle ***A-B-C-A*** loop in Fig. 3a encloses a flux of $\pi/\mathbf{3}$ ($-\pi/\mathbf{3}$) for the spin up (down) hole, for the example of twisting moiré or biaxial moiré.

The out-of-plane mirror symmetry dictates that $\varepsilon_B = \varepsilon_C$, and $t_1 = t_2$. Because of the higher barrier between ***B*** and ***C***, $t_3 < t_{1,2}$. The onsite energy $\varepsilon_A$ is larger than $\varepsilon_{B,C}$ in the scalar potential shown in Fig. 2a. In a sufficiently large moiré, where the hopping becomes exponentially small, $|t_{1,2,3}| \ll \varepsilon_A - \varepsilon_{B,C}$, we can adiabatically eliminate the $A$ sites. The resultant two-orbital tight-binding model consisting of the ***B*** and ***C*** sites becomes (Fig. 3b),

$$\hat{H}_{TB} = \sum_i \left(\varepsilon_B \hat{B}_l^\dagger \hat{B}_l + \varepsilon_C \hat{C}_l^\dagger \hat{C}_l\right) - \sum_{\langle l,m \rangle} \left(t_3 \hat{C}_l^\dagger \hat{B}_m + \textbf{h.c.}\right)$$
$$- \sum_{\langle\langle l,m \rangle\rangle} \left(t_B \exp\left(i\phi_B^{l,m}\right) \hat{B}_l^\dagger \hat{B}_m + t_C \exp\left(i\phi_C^{l,m}\right) \hat{C}_l^\dagger \hat{C}_m + \textbf{h.c.}\right)$$

where $t_B = \frac{t_1^2}{\varepsilon_A - \varepsilon_B}$, $t_C = \frac{t_2^2}{\varepsilon_A - \varepsilon_C}$, and $\langle\langle ... \rangle\rangle$ runs over next nearest-neighbor pairs. $\phi_B^{l,m}, \phi_C^{l,m} = \pm \frac{2\pi}{3}$, and the positive phase hopping directions are indicated by the arrows shown in Fig. 3b. This realizes the Kane-Mele model(32) and the Haldane model in each spin subspace(33), which explains the quantized spin Hall conductance found in mini-band calculations in the TMD homobilayer moiré (19, 25).

Fig. 3c shows the dispersions and the momentum-space Berry curvatures $\Omega^k$ of the three mini-bands, calculated with the parameters $\varepsilon_A - \varepsilon_{B,C} = \mathbf{5\ meV}$, $t_{1,2} = \mathbf{1\ meV}$ and $t_3 = \mathbf{0.5\ meV}$. The topological numbers of the three bands are $C = \frac{1}{2\pi} \mathbf{e}_z \cdot \int_{\text{mBZ}} \Omega^k d\mathbf{k} = \mathbf{-1}, \mathbf{+1}$ and $\mathbf{0}$. This three-band tight-binding model well reproduces the band dispersions, the $\Omega^k$ distributions, and topological numbers from the direct mini-band calculations in twisted MoTe$_2$ bilayer(19).

**Geometric scalar potential.** The off-diagonal Berry connection can become increasingly important with the decrease of the moiré period $b$. In the Hamiltonian in Eq. (4), $\mathbf{A}^{\pm\mp}$ plays two roles. First, it contributes $G(\mathbf{R}) \equiv \frac{\hbar^2}{2m^*} \mathbf{A}^{+-} \cdot \mathbf{A}^{-+}$ to the diagonal element of Hamiltonian. As a result, the scalar superlattice potential is corrected from $-E_\pm(\mathbf{R})$ to $-E_\pm(\mathbf{R}) + G(\mathbf{R})$. Second, it introduces a residue coupling between the layer-pseudospin branches $|+\rangle$ and $|-\rangle$, with the coupling form: $\left(i\frac{\partial}{\partial \mathbf{R}} + \mathbf{A}^{++}\right) \cdot \mathbf{A}^{+-} + \mathbf{A}^{+-} \cdot \left(i\frac{\partial}{\partial \mathbf{R}} + \mathbf{A}^{--}\right) \propto b^{-2}$. However, in the lower branch $|+\rangle$, low energy states are detuned from the branch $|-\rangle$ by a gap independent of $b$ (c.f. Fig. 2a), which quenches the off-diagonal effect except for very small moiré ($b \leq \mathbf{5}$ nm). Therefore, the low energy holes are well described by the effective Hamiltonian:

$$\hat{H}^{(+)} \approx \frac{\hbar^2}{2m^*}\left(i\frac{\partial}{\partial \mathbf{R}} + \mathbf{A}^{++}\right)^2 - E_+(\mathbf{R}) + G(\mathbf{R}). \quad (7)$$

$G(\mathbf{r})$ here is an additional energy due to the precession of the layer-pseudospin in the adiabatic transport(34), which repels electrons (holes) from the position where the pseudospin changes quickly. This effective potential of the geometric origin is known as the geometric scalar potential. In a homobilayer moiré, $G(\mathbf{R}) \propto b^{-2}$, increasing rapidly with the decrease of the moiré period $b$, and peaking between $B$ and $C$ points of the supercell (Fig. 4a).

Fig. 4c plots the overall scalar potential at different moiré periods $b$. The correction by the geometric contribution becomes significant at $b \leq 20$ nm. The local out-of-plane mirror symmetry at $A$ dictates $G(\mathbf{R})$ vanishes at this location. The geometric scalar potential therefore pushes $\varepsilon_{B,C}$ towards $\varepsilon_A$ (see Fig. 4c), so the $A$ orbital can no longer be perturbatively eliminated in a small moiré. From the three-orbital tight-binding calculation using Eq. (6), we find a topological phase transition occurs when $t_3$ is increased to $(\varepsilon_A - \varepsilon_{B,C})/3$. In Fig. 3d, we show the results with parameters $\varepsilon_A = \varepsilon_{B,C} = 0$, $t_1 = t_2 = 5$ meV and $t_3 = 2$ meV, where the topological numbers of the three bands become $\mathcal{C} = -1, -1$ and $+2$, respectively. The tight-binding calculations shown in Fig. 3c and 3d reproduce the distinct moiré band dispersions and topologies in MoTe$_2$ bilayers at twist angles 1.2° and 2° respectively. The change of the energy landscape by the geometric scalar potential accounts for the topological phase transition as a function of the twist angle.

**Quantized switch of magnetic flux by interlayer electric bias.** An interlayer bias can add a tunable term $\frac{1}{2}e\mathcal{E}d\hat{\sigma}_z$ to $\hat{H}_\nu(\mathbf{R})$, where $d$ is the interlayer distance, and $\mathcal{E}$ is the perpendicular electric field. The effect is to replace $\Delta(\mathbf{R})$ by $\Delta_\mathcal{E}(\mathbf{R}) \equiv \Delta(\mathbf{R}) - \frac{1}{2}e\mathcal{E}d$, which changes the profiles of both the potential landscape $E_\pm(\mathbf{R})$ and the magnetic field $\mathcal{B}^\pm(\mathbf{R})$. The electric bias can thus be exploited to continuously tune the moiré magnetic field and the flux superlattice.

The lower panel of Fig. 5a shows the magnetic field distribution in the pseudospin branch $|+\rangle$ under an interlayer bias $\mathcal{E}d = 0.026$ V in the twisting moiré. The magnetic field distribution is significantly changed compared to the zero-bias distribution shown in Fig. 5a upper panel, having been pushed from the $B$-$C$ center towards the $C$ point where the difference between $E_+(\mathbf{R})$ and $E_-(\mathbf{R})$ is minimal. The magnetic flux per supercell remains $2\pi$. With further increase of bias to the critical value $\mathcal{E}d = 2\Delta_0/e = 0.044$ V, $\Delta_\mathcal{E}$ crosses $0$ at $C$ point where $h(\mathbf{R})$ also vanishes (c.f. Fig. 2b). Therefore the gap between $E_+(\mathbf{R})$ and $E_-(\mathbf{R})$ closes, whereupon a topological transition of the layer pseudospin texture occurs (Fig. 5b). Fig. 5c shows the magnetic field distribution after this transition ($\mathcal{E}d = 0.06$ V), where the magnetic flux per supercell becomes $0$.

**Discussions**

Massive Dirac Fermion features an orbital magnetic moment $\mathbf{M} = \frac{\hbar^2}{2m^*}\mathbf{e}_z$ arising from the momentum-space Berry phase effect(35), which leads to Zeeman shifts of the band edges of monolayer TMDs in an external magnetic field(36-38). It is interesting to ask whether the emergent moiré magnetic field $\mathcal{B}^+$, as a manifestation of the real-space Berry phase, can couple to the orbital magnetic moment as well. In the perturbative expansion from Eq. (2) to Eq. (7), we can separate two contributions in the second order perturbative correction: $\widehat{H}_+ = -E_+(\mathbf{R}) + E_s^{(2)} + E_o^{(2)}$,

$$E_s^{(2)} = \frac{\hbar^2 v_D^2}{E_g}(i\partial_- + \mathbf{e}_- \cdot \mathbf{A}^{++})(i\partial_+ + \mathbf{e}_+ \cdot \mathbf{A}^{++}),$$

$$E_o^{(2)} = \frac{\hbar^2 v_D^2}{E_g}(\mathbf{e}_- \cdot \mathbf{A}^{+-})(\mathbf{e}_+ \cdot \mathbf{A}^{-+}).$$

$E_s^{(2)}$ is from the coupling between $|+\rangle = \widehat{U}(\mathbf{R})|t\rangle_v$ and the conduction state of the *same* pseudospin orientation $|c_+\rangle = \widehat{U}(\mathbf{R})|t\rangle_c$ (see Fig. 4d), which can be rewritten as: $E_s^{(2)} = \frac{\hbar^2 v_D^2}{E_g}\left(i\frac{\partial}{\partial \mathbf{R}} + \mathbf{A}^{++}\right)^2 - \mathcal{B}^+ \cdot \mathbf{M}$. In addition to the kinetic energy term, it does contain the coupling of the orbital magnetic moment to the moiré magnetic field $\mathcal{B}^+$, which contributes to the scalar potential landscape (Fig. 4b). Because of the non-Abelian nature of the Berry phase effect here, there is also a contribution $E_o^{(2)}$ from the coupling between $|+\rangle$ and the conduction state of *opposite* pseudospin $|c_-\rangle = \widehat{U}(\mathbf{R})|b\rangle_c$. The Zeeman energy partially cancels with $E_o^{(2)}$, with the net effect being the geometric scalar potential $G(\mathbf{R})$.

In contrast to the twisting moiré, the moiré introduced by a biaxial or uniaxial strain allows engineering moiré magnetic field and flux superlattice through mechanical controls. In a rotationally aligned homobilayer, the application of a relative strain $\eta$ between the two layers, e.g. through substrate, can create a moiré pattern with period $b = a/\eta$. A modest tensile strain can thus tune $b$ over several orders of magnitude, where the magnetic field scales as $b^{-2}$. This homobilayer moiré points to unprecedented opportunities to explore physics in an ultra-high magnetic field that can be tailored on the nanoscale by both electrical and mechanical means.

It is also interesting to compare with twisted bilayer graphene, where the interplay of sublattice pseudospin and layer pseudospin leads to a more complex gauge structure for the massless Dirac fermion(28, 39). Only at the AA stacking locales(39) can the gauge structure in a graphene moiré be simplified to a form equivalent to that of a magnetic field $\propto b^{-1}$ (in contrast to the $b^{-2}$ scaling underlying the $b$-independent flux quantization here). This gives rise to pseudo Landau levels of the massless Dirac fermion at the AA locales, which can be connected to the flat mini-bands in the bilayer graphene at the magic twist angle(28, 39). For massive particles here, the strongly inhomogeneous magnetic field profile in a moiré supercell of exactly quantized flux points to manifestations of ultra-high magnetic fields that are remarkably different from the well-studied Landau-level physics.

The flux lattices and quantum spin Hall effects need to be explored at low doping in a relatively small moiré, since the topological gap (determined by hopping integrals between superlattice sites) decreases exponentially with $b$. The other limit, i.e. relatively high doping in a large moiré, is equally interesting to explore, where the nanoscale patterned magnetic field in the associated potential landscape (Fig. 5) points to a new realm of magneto transport. Electrostatic doping and interlayer bias control of the moiré magnetic flux can both be exploited to tune the magnetic field experienced by an electron gas of a periodically varying density. With the valley-spin dependent sign, the moiré magnetic field also leads to the valley and spin Hall effects in such a metallic regime. Remarkable gate tunability of the valley-spin Hall conductance can be expected from both the bias dependence and carrier density dependence of moiré magnetic flux experienced by the carriers. The non-local transport measurement in Hall bar geometry can be used to detect such effects. We also expect other important manifestations of the moiré magnetic field when an external magnetic field is applied. With carriers experiencing the sum or difference of the moiré and applied magnetic fields at the two valleys, their response quantified with respect to the external magnetic field, e.g. magnetoresistance, can be significantly changed by the moiré field.

Our study here focuses on **±K**-valley carriers only. In certain TMD homobilayers, the **±K**-valley is still the band edge: for example, the valence band edge of bilayer $WSe_2$(40) and conduction and valence band edges of bilayer $MoTe_2$(41). In other compounds, **Γ** pocket of holes and/or **Λ** pockets of electrons become relevant. In such cases, the quantum spin Hall effect may not be possible to observe, as the topological gap overlaps with the Fermi sea at **Γ** or **Λ**. Nevertheless, other important manifestations of the moiré magnetic field (like the valley-spin Hall effect in the metallic regime) can still be explored for the carriers at **±K**-valleys, even if they are not the band edge.

**Methods:** DFT calculations of the bilayer $MoSe_2$ structure, which is modeled by a slab, were performed using the Vienna ab initio Simulation Package(42). To avoid artificial interactions between the polar slabs when the bilayer deviates from the AA stacking, two such slabs, oppositely oriented with mirror symmetry, are placed in each cell, which is separated from its periodic images by 20 Å vacuum regions(43). The cutoff energies for the plane wave basis set used to expand the Kohn-Sham orbitals are 400 eV for all calculations. The exchange correlation functional is approximated by the generalized gradient approximation as parametrized by Perdew, Burke and Ernzerhof(44). The 2D Brillouin zone is sampled by a 30x30x1 Monkhorst-Pack mesh. Van der Waals dispersion forces between the adsorbate and the substrate were accounted for through the optB88-vdW functional by using the vdW-DF method developed by Klimeš *et al.* (45).

**Acknowledgments:** We acknowledge support by the Research Grants Council of Hong Kong (17306819), Croucher Foundation, and Seed Funding for Strategic Interdisciplinary Research Scheme of HKU.

**Table**

|  | Twisting ($\theta$) | Biaxial strain ($\eta$) | Uniaxial strain ($\eta$) |
|---|---|---|---|
| moiré period ($b$) | $b = a/\theta$ | $b = a/\eta$ | $b = a/\eta$ |
| supercell primitive vectors | $\mathbf{b}_1 = -\frac{b}{2}\mathbf{e}_x + \frac{\sqrt{3}b}{2}\mathbf{e}_y$ $\mathbf{b}_2 = \frac{b}{2}\mathbf{e}_x + \frac{\sqrt{3}b}{2}\mathbf{e}_y$ | $\mathbf{b}_1 = \frac{\sqrt{3}b}{2}\mathbf{e}_x + \frac{b}{2}\mathbf{e}_y$ $\mathbf{b}_2 = \frac{\sqrt{3}b}{2}\mathbf{e}_x - \frac{b}{2}\mathbf{e}_y$ | $\mathbf{b}_1 = \frac{\sqrt{3}b}{2}\mathbf{e}_x - \frac{b}{2}\mathbf{e}_y$ $\mathbf{b}_2 = \frac{\sqrt{3}b}{2}\mathbf{e}_x + \frac{b}{2}\mathbf{e}_y$ |
| local registry ($\mathbf{r}$) as a function of position ($\mathbf{R}$) | $\mathbf{r}_x = \theta R_y$ $\mathbf{r}_y = -\theta R_x$ | $\mathbf{r} = \eta \mathbf{R}$ | $\mathbf{r}_x = \eta R_x$ $\mathbf{r}_y = -\eta R_y$ |
| magnetic flux per supercell — $\|\mathcal{E}d\| < 2\Delta_0/e$ | $2\pi$ | $2\pi$ | $-2\pi$ |
| magnetic flux per supercell — $\|\mathcal{E}d\| > 2\Delta_0/e$ | 0 | 0 | 0 |

**Table 1 Comparison of flux superlattices in three types of homobilayer moiré.** One monolayer has the primitive lattice vectors $\mathbf{a}_1 = \frac{\sqrt{3}a}{2}\mathbf{e}_x + \frac{a}{2}\mathbf{e}_y$, $\mathbf{a}_2 = \frac{\sqrt{3}a}{2}\mathbf{e}_x - \frac{a}{2}\mathbf{e}_y$, and the other layer is rotated by a twisting angle $\theta$, or subject to a biaxial or uniaxial tensile strain respectively. The uniaxial strain is applied along the armchair or zigzag direction, with the area of the strained unit cell conserved. The magnetic flux per moiré supercell is given for the spin-up carrier ($-\mathbf{K}$-valley). The local registry is parameterized by the in-plane displacement $\mathbf{r}$ between a near-neighbor pair of atoms from the opposite layers (c.f. Fig. 1 inset). $\mathcal{E}d$ is the interlayer electric bias, and its critical value $2\Delta_0/e$ = **0.044 V**.

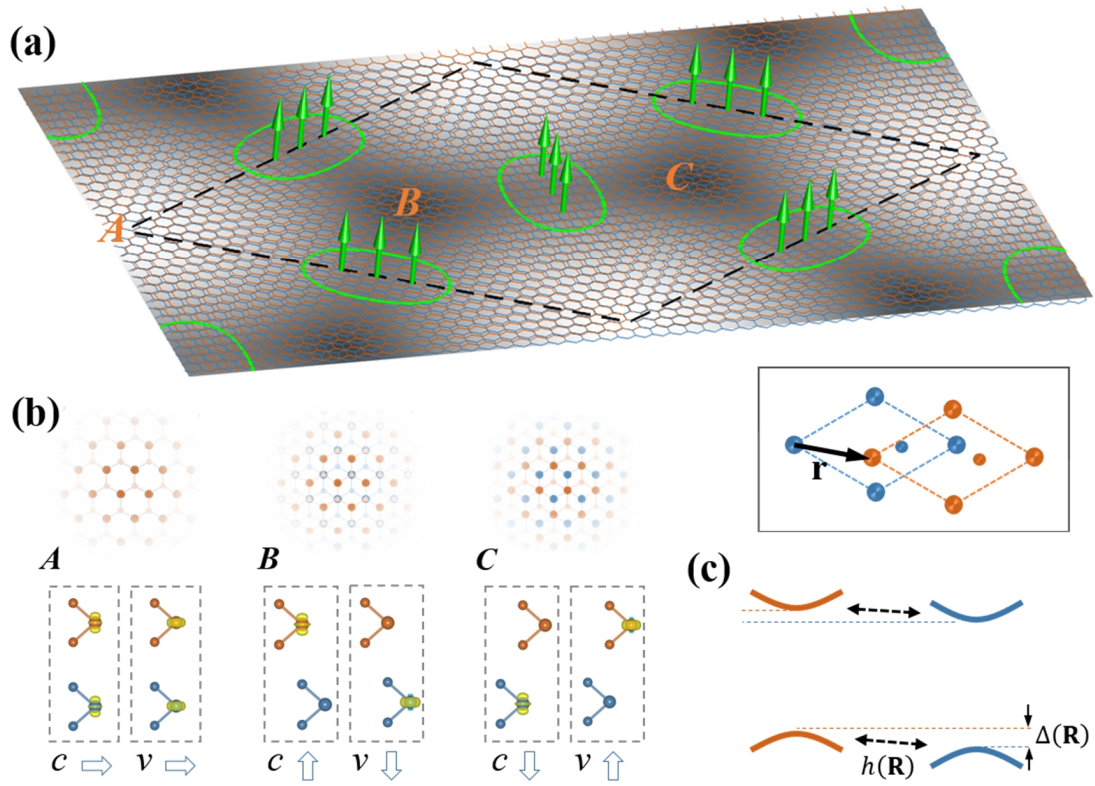

**Figure 1. Moiré pattern as a real-space texture for Berry phase.** (a) Schematic of a homobilayer moiré, and periodic magnetic flux (green arrows) from the real-space Berry phase. The dashed rhombus denotes a supercell. (b) Upper panel: atomic registries at three high-symmetry locales $A$, $B$ and $C$. Lower panel: corresponding layer distributions of conduction ($c$) and valence ($v$) band-edge carriers (yellow isosurfaces), with the arrows indicating the layer pseudospin orientations. (c) A schematic of the coupled massive Dirac cones from the two layers. The interlayer coupling parameters ($h, \Delta$) depend on the interlayer registry characterized by the displacement **r** between a near-neighbor pair of atoms from two layers (inset); in a moiré, they are functions of location **R**.

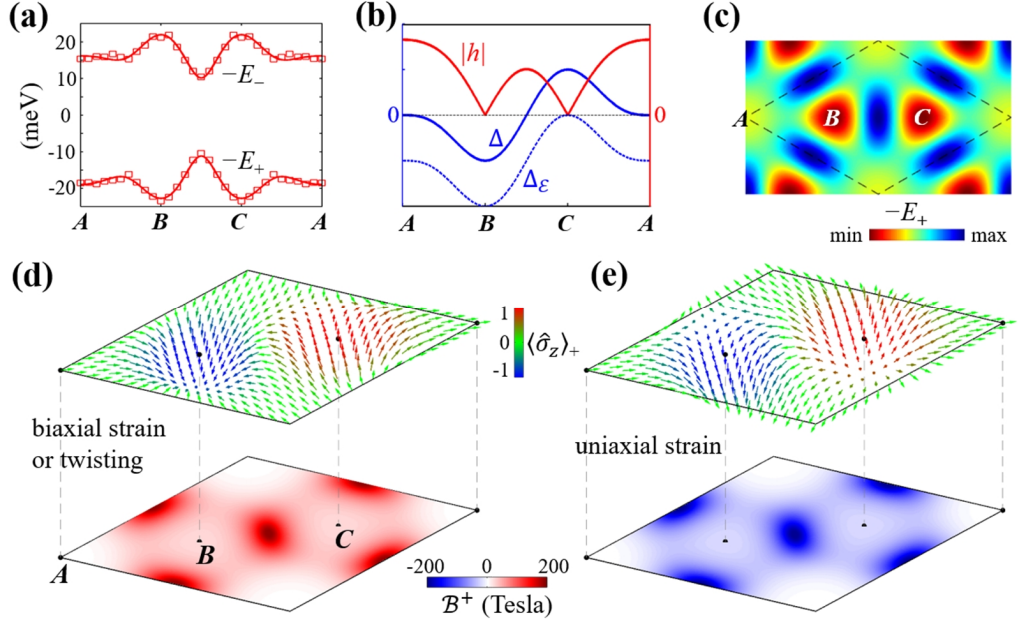

**Figure 2. Layer-pseudospin texture and moiré magnetic field. (a)** The symbols are from the *ab initio* calculations of the split valence band edges $E_+$ and $E_-$ in MoSe$_2$ homobilayers of various interlayer registries. The curves are fitted using Eq. (5) with the parameters $\delta_0 = $ **1 meV**, $\Delta_0 = $ **22.3 meV**, $h_0 = $ **7.1 meV** and $h_1 = -$**1.2 meV**. **(b)** The corresponding interlayer coupling parameters as functions of position **R** in moiré. $\Delta_{\mathcal{E}} \equiv \Delta - \frac{1}{2}e\mathcal{E}d$ at the critical bias value $\mathcal{E}d = $ **2**$\Delta_0/e$ is shown as the dotted curve. **(c)** Color map showing the scalar potential landscape $-E_+$ in a moiré supercell, where the high-symmetry locales $A$, $B$ and $C$ are energy minima. **(d)** and **(e)** Upper panel: layer pseudospin texture in the lower energy pseudospin branch $|+\rangle$. The arrows show $\langle\boldsymbol{\sigma}\rangle_+$ at different positions, and the color codes its $z$-component. Lower panel: magnetic field $\mathcal{B}^+$ in a moiré of period $b = $ **10** nm. The peak value reaches 200 T. (d) is for a moiré induced by a twisting or biaxial strain, and (e) is for a moiré induced by a uniaxial strain (c.f. Table 1).

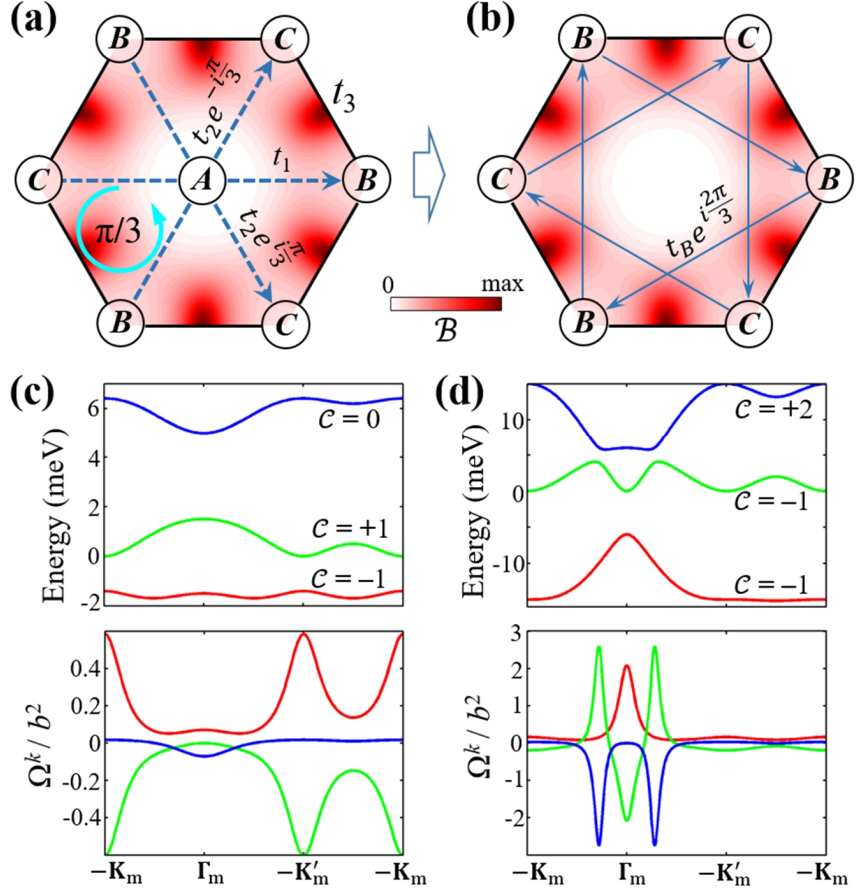

**Figure 3 Flux superlattice for quantum spin Hall effect. (a)** The color map shows the moiré magnetic field $\mathcal{B}^+$. The lines illustrate hopping between trapping sites $A$, $B$ and $C$. The phases for the hopping direction indicated by the dashed arrows are shown. **(b)** The effective 2-orbital model in a long-period moiré where the high energy trapping sites $A$ are perturbatively eliminated. The positive phase directions of the next-nearest neighbor hopping are indicated by the arrows. **(c)** and **(d)** Band dispersions and momentum-space Berry curvature $\Omega^k$ of the mini-bands calculated using the 3-orbital model in (a). For (c), the parameters are $\varepsilon_A = 5$ **meV**, $\varepsilon_B = \varepsilon_C = 0$, $t_1 = t_2 = 1$ **meV** and $t_3 = 0.5$ **meV**, and the three mini-bands have the Chern numbers of $\mathcal{C} = -1$, $+1$ and $0$, respectively. For (d), $\varepsilon_A = \varepsilon_B = \varepsilon_C = 0$, $t_1 = t_2 = 5$ **meV** and $t_3 = 2$ **meV**, and the band Chern numbers become $\mathcal{C} = -1, -1$ and $+2$, respectively.

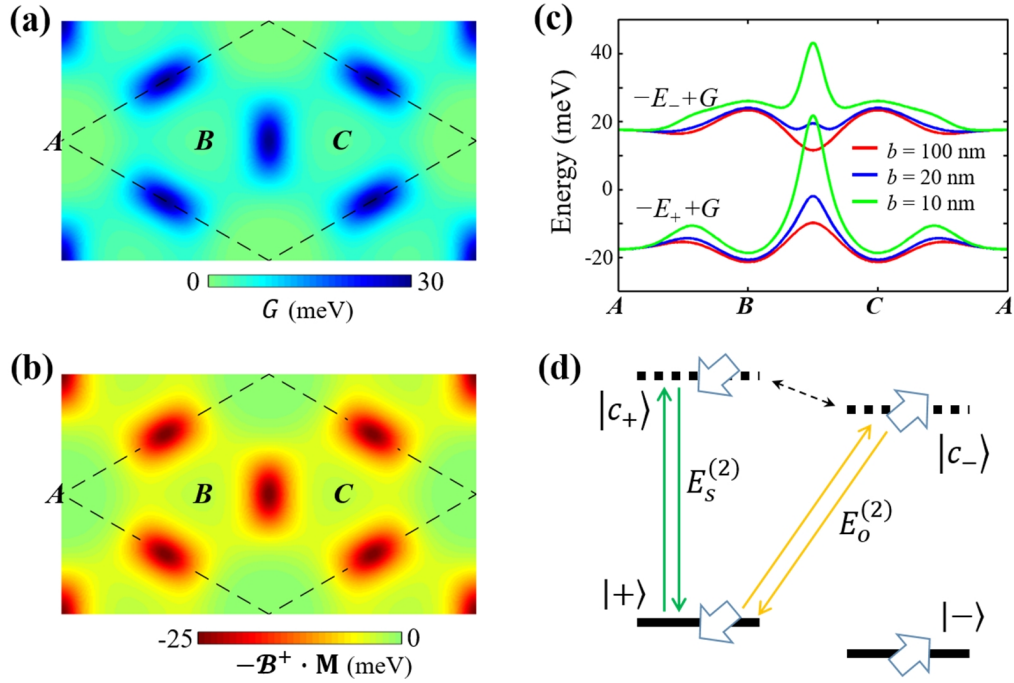

**Figure 4. Geometric scalar potential from the non-Abelian Berry connection.** (**a**) The geometric scalar potential $G$ for a moiré of period $b = 10$ nm. (**b**) The Zeeman energy $-\mathcal{B}^+ \cdot \mathbf{M}$, where $\mathbf{M}$ is the orbital magnetic moment of massive Dirac Fermion from the momentum-space Berry phase. (**c**) The overall scalar potential $-E_\pm + G$ in the two pseudospin branches at different moiré period. (**d**) The schematic level scheme for the non-Abelian Berry phase effect in a homobilayer moiré. $|+\rangle$ is coupled to the conduction state of the same pseudospin $|c_+\rangle$, and the second order perturbative correction $E_s^{(2)}$ accounts the moiré magnetic field $\mathcal{B}^+$ effects, including the Zeeman energy shown in (b). The coupling of $|+\rangle$ to conduction state of the opposite pseudospin $|c_-\rangle$ leads to another correction $E_o^{(2)}$, and its sum with the Zeeman energy term is the overall geometric scalar potential $G$ shown in (a).

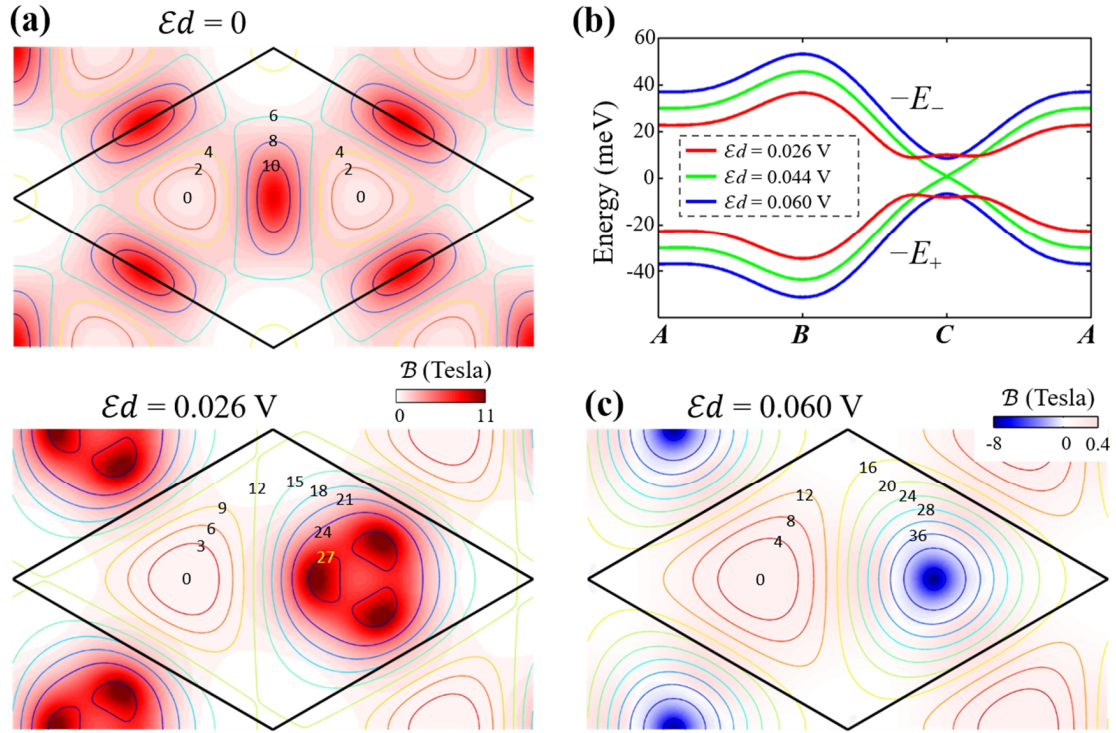

**Figure 5. Tuning of magnetic field profile and quantized change of magnetic flux by interlayer bias.** **(a)** The magnetic field profile and energy contour in the lower pseudospin branch $|+\rangle$ under an interlayer bias $\mathcal{E}d = 0.026$ V (lower panel), compared to the zero-bias profile (upper panel). The moiré period $b = 50$ nm. The flux per supercell is $2\pi$ in both cases. **(b)** Scalar superlattice potentials in the two pseudospin branches at three bias values. At the critical bias **0.044 V**, $-E_+$ and $-E_-$ touch at $C$, whereupon a topological transition of the layer-pseudospin texture occurs. **(c)** The magnetic field distribution at a bias of **0.06V**, where the flux per supercell is 0.